\documentclass{moriond}

\def\Journal#1#2#3#4{{#1} {\bf #2}, #3 (#4)}

\def\be{\begin{equation}}
\def\ee{\end{equation}}
\def\bea{\begin{eqnarray}}
\def\eea{\end{eqnarray}}

\newcommand{\Photo}{}

\begin{document}
\vspace*{4cm}
\title{Overview of radio experiments for UHE cosmic particles detection}

\author{ Simon Chiche }
\address{Sorbonne Universit\'{e}, CNRS, UMR 7095, Institut d'Astrophysique de Paris, 98 bis bd Arago, 75014 Paris, France}

\author{Valentin Decoene}

\address{Department of Physics, The Pennsylvania State University, USA, University Park 16802 PA}
\address{Center for Multimessenger Astrophysics, Institute for Gravitation and the Cosmos, The Pennsylvania State University, USA, University Park 16802 PA}

\maketitle\abstracts{
Radio-detection is a mature technique that has gained large momentum over the past decades. Its physical detection principle is mainly driven by the electromagnetic part of the shower, and is therefore not too sensitive to uncertainties on hadronic interactions. Furthermore its technical detection principle allows for a 100\% duty cycle, and large surface coverage thanks to the low cost of antennas. 
Various detection methods of UHE particles now rely on the radio signal as main observable. For instance, ground based experiments such as AERA on the Pierre Auger Observatory or LOFAR detect the radio emission from air-showers induced by high-energy particles in the atmosphere; in-ice experiment such as ARA, IceCube, or ARIANNA benefits from a detection in denser media which reduces the interaction lengths; finally, balloon experiments such as ANITA allow for very sensitive UHE neutrino detection with only a few antennas. Radio-detection is now focused on building increasingly large-scale radio experiments to enhance the detector sensitivity and address the low fluxes at UHE. In this proceeding we give an  overview of the past, current and future experiments for the detection of UHE cosmic particles using the radio technique in air (AERA, Auger-Prime, GRAND), in balloon (ANITA, PUEO) or in other media (IceCube-Gen2, BEACON, RNO-G).}

\section{Motivations for Radio-detection of ultra-high energy astroparticles}

 Over the last decade, high energy astroparticles have been observed in coincidence with several high energy events such as TXS0506+056~\cite{TXS} and AT2019dsg~\cite{AT2019}. Yet, more than 50 years after the first observation of ultra-high energy cosmic rays (UHECRs), their origin is still unknown. Towards the understanding of this long-standing mystery increasing experimental efforts are made to detect cosmic-rays and their secondary messengers (gamma rays and neutrinos) at the highest energies despite the very low incoming flux. This  would allow to not only constrain UHECRs sources but also to probe the most powerful sources in the universe in the advent of the multi-messenger era.

\begin{figure}[tb]
\centering 
\includegraphics[width=0.99\columnwidth]{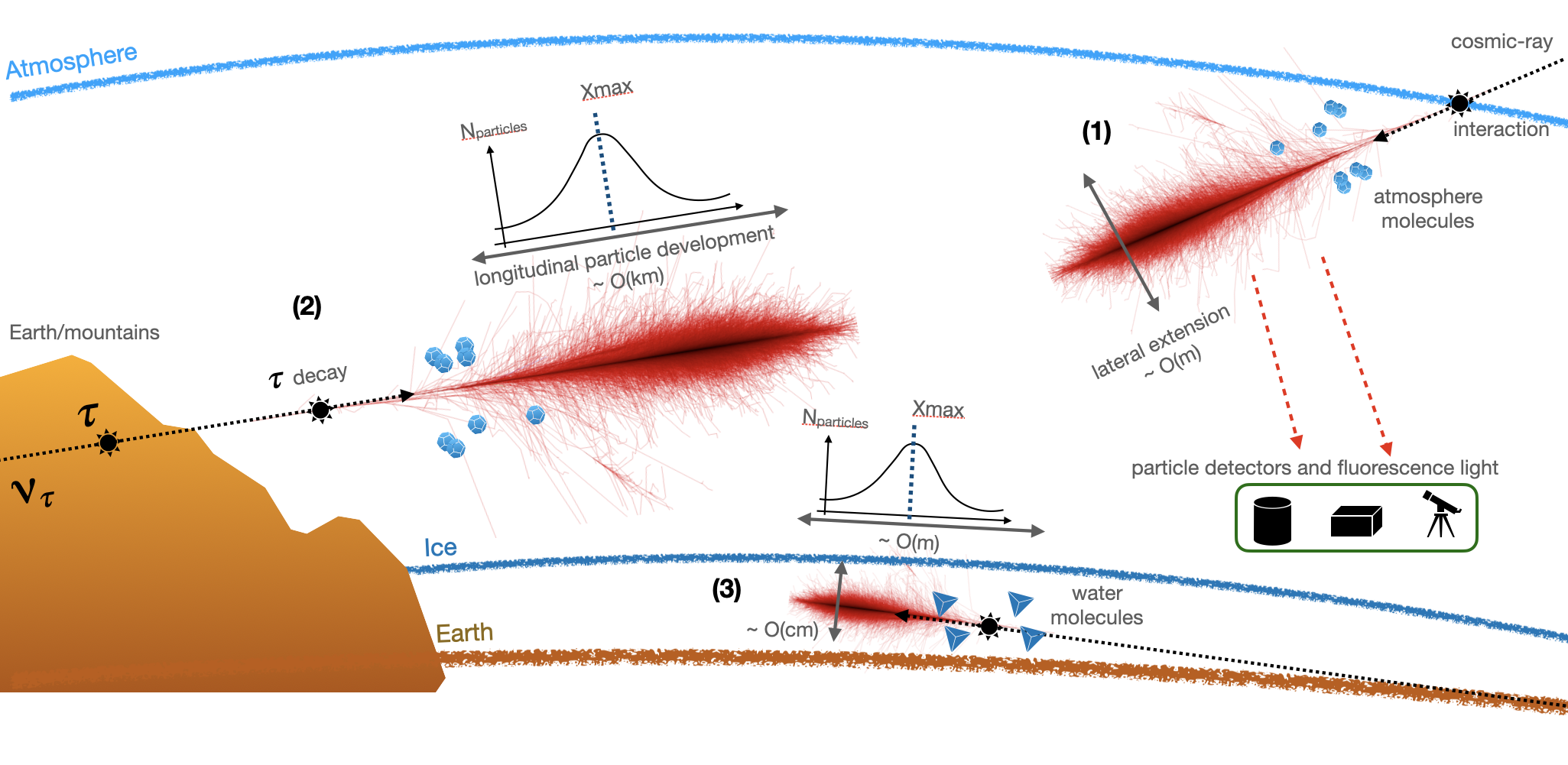}
\caption{Sketch of UHE astroparticles detection principle for (1) in-air showers, (2) tau-neutrino induced showers and (3) in-ice showers. The typical longitudinal profile is represented atop of the showers.}\label{fig:Radio_detection}
\end{figure}

The detection of high-energy astroparticles relies on the signal from the particle cascades they induce while arriving on Earth as illustrated in Fig.~\ref{fig:Radio_detection}. Typically the interaction of a cosmic-ray with air atoms from the atmosphere creates an extensive air-shower that propagates over 10 to 100 of kilometers and emits Cerenkov radiation, fluorescence light and electromagnetic radiation detectable on Earth alongside with the particles reaching the ground~\cite{Schr_der_2017}.
Similar showers can also be induced by tau-neutrinos arriving with Earth-skimming trajectories going through a dense medium such as a mountain or Earth's surface. In these media, the neutrino can interact to give a tau particle, which then, can escape to decay in the atmosphere inducing an air-shower. Finally, particle cascades can develop in denser media such as ice or water, where due to the higher density, the showers extent over smaller distances, of the order of a few meters for the longitudinal profile and of a few centimeters for the lateral one. Similar to air-showers these, in ice showers also emit Cerenkov radiation and electromagnetic-waves.

The radio emission from particle showers has been extensively studied over the past decades. 
For in air showers, the radiation results from two main mechanisms, with an intensity peaking in the $ \rm MHz$ regime. (1) The geomagnetic emission: the deflection of the lightest charged particles in the shower, i.e., positrons and electrons in opposite directions induce a current varying in time as the particle content in the shower varies over time leading to a radio signal polarized along the $-\mathbf{v} \times \mathbf{B}$ direction (with $\mathbf{B}$ being the direction of the magnetic field and $\mathbf{v}$ the direction of the shower). (2) The charge-excess or Askaryan emission: while the shower propagates, electrons from air-atoms are struck by high energy shower particles and then travel along with the shower front. This combined with positron annihilation leads to a build up of a net negative charge in the shower front inducing a signal radially polarized in a plane perpendicular to the shower axis. The geomagnetic emission is dominant in air where the charge-excess account for only $\sim 10\%$ of the signal but it is negligible in ice where the charge-excess is much stronger and corresponds to the dominant emission.
In addition to that, the radio signal undergoes a geometrical time compression phenomenon called Cerenkov effect. It is due to the relativistic speed of the particles which confine the emission in a cone with a typical aperture angle of $1^{\circ}$ for an air-shower and of roughly $40^{\circ}$ to $60^{\circ}$ for in ice showers.

\section{First generation of radio experiments}
It was proved in 1965 that air-showers emit radio waves~\cite{Huege_2016}, yet it is only in the 2000s that radio detection really took off, mainly due to the improvements in digital signal processing and motivated by an expected duty cycle of $100\%$.
\subsection{CODALEMA and LOPES}
\begin{figure*}[tb]
\centering 
\includegraphics[width=0.44\columnwidth]{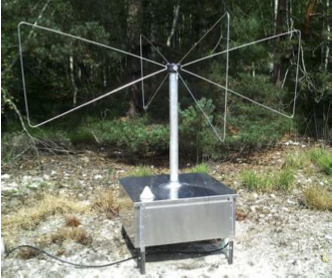}
\includegraphics[width=0.54\columnwidth]{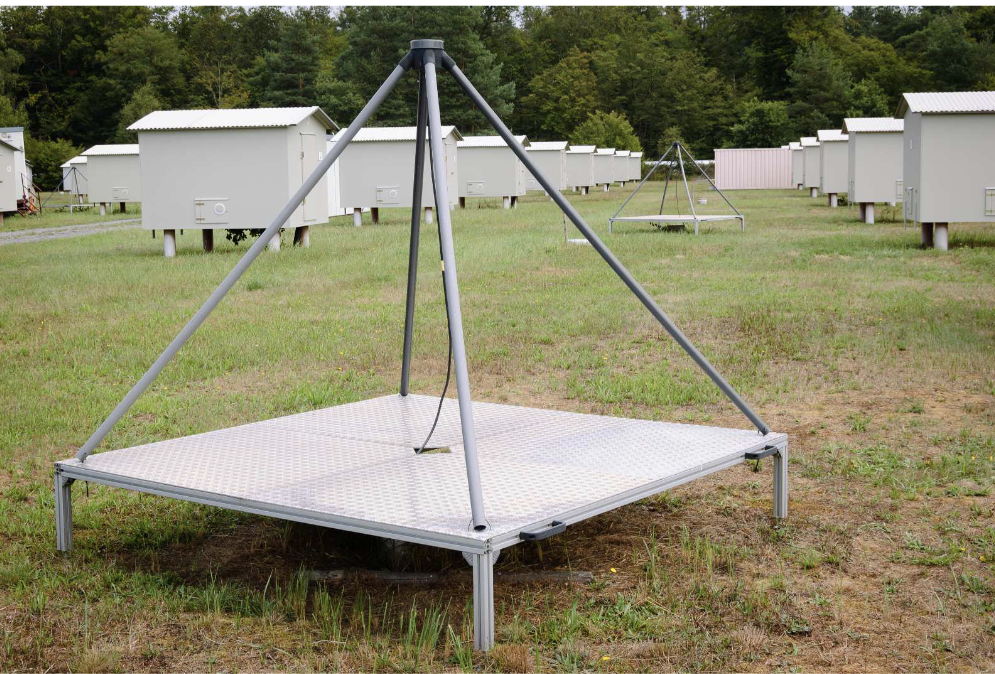}
\caption{{\it Left} CODALEMA butterfly tripole antenna from~\protect\cite{CODALEMA_antenna} and {\it right} LOPES inverted v-shape dipole antenna from~\protect\cite{Apel_2021}.}
\label{fig:radio_antenna}
\end{figure*}

 This emergence of radio-detection as a promising technique was led by 2 pioneering experiments,  \textbf{CODALEMA}~\cite{Ardouin_2005,Ardouin_2009,Charrier_2019c} and \textbf{LOPES}~\cite{Huege_2012,Apel_2014,Apel_2021} which aimed at probing that radio-detection of cosmic-ray induced showers in the atmosphere was feasible. These experiments relied on the fact that radio waves emitted by particle cascades travel in air with almost no absorption and could be detected at ground level with radio-antennas. Radio-detection was however impeded by the numerous anthropogenic radio emissions from various sources such as the FM band, airplanes or satellites.  
CODALEMA, the Cosmic-ray Detection Array with Logarithmic Electro-magnetic Antennas is an experiment initiated in the Nancay radio-astronomy station in 2003. It combined a sparse array of 57 autonomous radio antennas (Fig.~\ref{fig:radio_antenna}, left panel) detecting signals in the 20-200 MHz band with a compact array of cabled antennas triggered by 13 scintillators over $1$\,km$^2$. LOPES, the LOFAR Prototype Station, is a radio experiment  made of 30 LOFAR prototype antennas (Fig.~\ref{fig:radio_antenna}, right panel) that ran between 2003 and 2013. It used interferometry triggered by the KASKADE-Grande experiment in the $40-80$ MHz band. Even though the layout was set in the radio loud environment of the Karlsruhe Institute of Technology, LOPES showed that a radio detection with energy and angular resolution comparable to particle array performances was achievable. It also allowed the radio community to better understand the radio-emission mechanisms, highlighting the clear dependency of the signal with the geomagnetic angle. 

\subsection{RICE and AURA}

 In parallel to the in-air pioneering experiments, two detector concepts were investigated for the in-ice detection of neutrino-induced in-ice showers.
These investigations were motivated by the fact that the ice cap offers a gigantic interaction volume for neutrinos, making possible, in principle, a gigatonne sized detector with only few antennas in a relatively radio quiet environment.
These two detector concepts called \textbf{RICE}, Radio Ice Cherenkov Experiment~\cite{Kravchenko_2003} (1995-2005) and \textbf{AURA}, Antarctic Under-ice Radio Array~\cite{Landsman_2009} (2003-2009), aimed at deploying sub-surface and deep-ice antennas respectively.
RICE was co-located with the AMANDA prototype and was composed of $20$ half-wave antennas ($200-1000$\,MHz), in a $200\times200$\,m$^{2}$ layout between $100-300$\,m depth.
AURA was deployed with IceCube in a layout of $6$ clusters of $4$ antennas ($100-450$\,MHz) between $300-1300$\,m.
Sub-surface arrays while being obviously more convenient for the deployment faces more refraction and ray-bending effects than deep-ice ones, where the ice is more cold and more stable.
One of the big advantage of the in-ice radio detection is that the ice provides a denser interaction medium than air for high energy neutrinos.
However, the radio attenuation length in ice is of only $\sim1$\,km on average (a factor thousand lower than in the atmosphere!). In addition the emission is only coherent along the Cherenkov cone, which because of the refractivity of the ice is frequency dependant and much larger than in the air ($\sim 55^{\circ}$ in the $100$s of MHz regime).
Furthermore the variations of refractive index with depth leads to ray-bending and refraction effects, which need to be taken into account for the reconstruction and the interpretation of the data.

Nevertheless, RICE and AURA managed to demonstrated the feasibility of this technique, opening the path to future in-ice experiments, still in operation today for some of them.

\section{Second generation of radio experiments}

After the validation of the radio-technique by the first generation of experiments, the second generation aimed at using this radio technique in order to measure the cosmic-ray fluxes at the highest energy (above $\rm EeV$) and to characterize the cosmic-ray composition at these energies. In terms of experimental goals this translated into large scale deployments to increase effective layout areas and into developments of more performing triggers to increase the detection efficiency. These objectives were at least partly achieved by LOFAR, ANITA, AERA and TREND.

\subsection{In air detection: LOFAR, TREND and AERA}

\textbf{LOFAR}~\cite{Nelles_2015,Schellart_2014,Corstanje_2015}, the LOw Frequency ARray is a  large scale radio interferometer, with more than 40 stations of $\sim100$ antennas each in Netherlands and across several European locations for astronomical observations. The antennas are divided in two types, LBA dipoles operating in the $30-80\,$MHz band and HBA tiles covering the $110-240\,$MHz band. The array relies on beamforming techniques to combine the phased signal of individual antennas and point specific locations in the sky. Air-shower detection is performed using a triggering from the LORA (LOfar Radboud Air-shower array) particle array with $\sim300$\,m diameter. LOFAR achieved high precision measurements of air-shower with a time resolution of a few ns, allowing to reach an $X_{\rm max}$ resolution of $\sim 20\,\rm g.cm^{-2}$ and to probe physics of the radio emission mechanism at an unprecedented level. 

\vspace{0.2cm}
\noindent \textbf{TREND}~\cite{Charrier_2019} the Tianshan Radio-array for Neutrino Detection was an experiment deployed in a remote valley in the Tianshan moutains in western China with $\sim 50$ antennas over $1.1$\,km$^{2}$ and ran from 2009 to 2013. It aimed at probing air-shower autonomous radio-detection, i.e., detection of air-showers with radio antennas only and no external trigger. It detected 730 Million of events over 314 days. Among them
564 were identified as air-shower candidates. The background contamination was estimated at 20\% in this final subset of candidates. Furthermore comparing the detected number of events with the expected one, lead to an intrinsic detector efficiency of 32\%. TREND was a prototype experiment deployed on the radio quiet site of the 21 CMA experiment, but still it was the first experiment to show that autonomous radio detection of air-showers was feasible. Hence thanks to self-triggering, TREND has become the seed of next generation large-scale experiments as GRAND.  

\begin{figure*}[tb]
\centering 
\includegraphics[width=0.56\columnwidth]{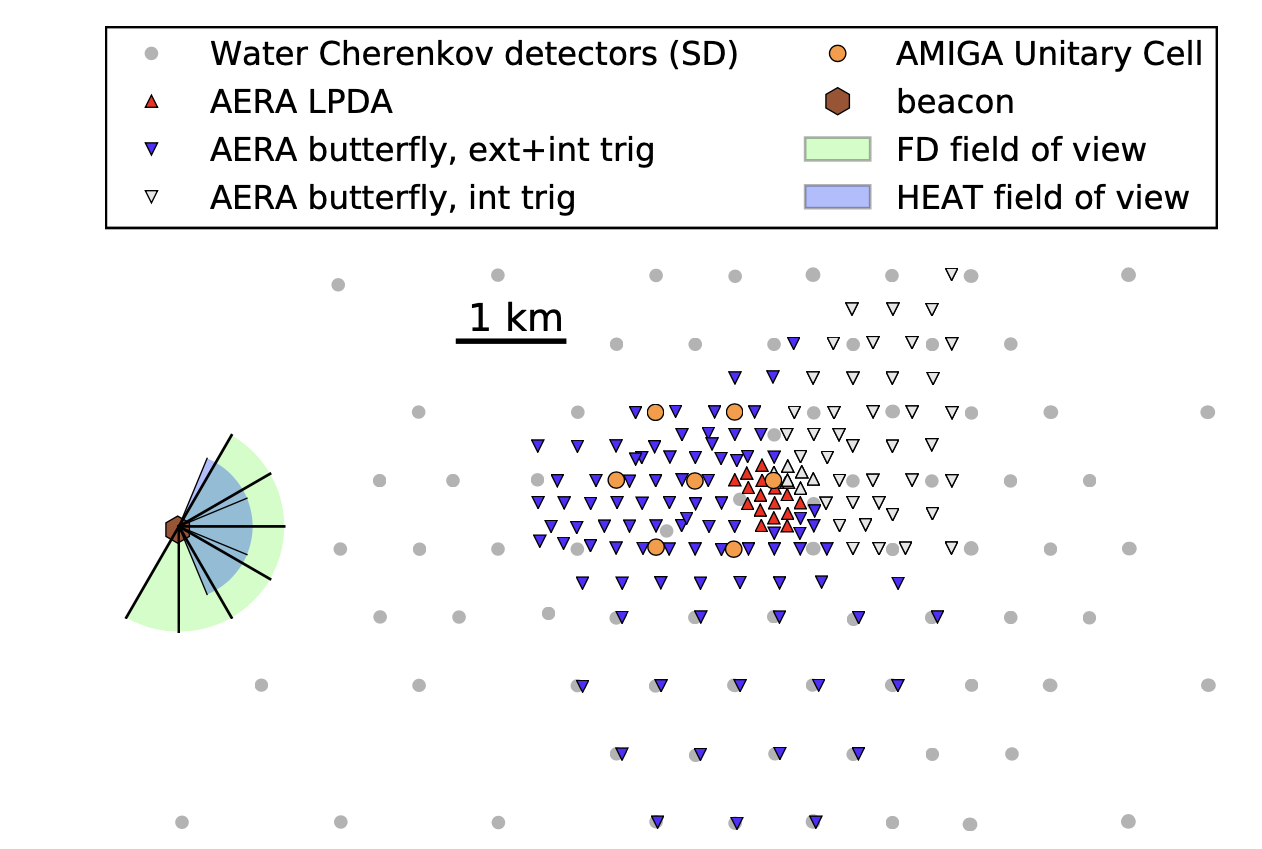}
\includegraphics[width=0.42\columnwidth]{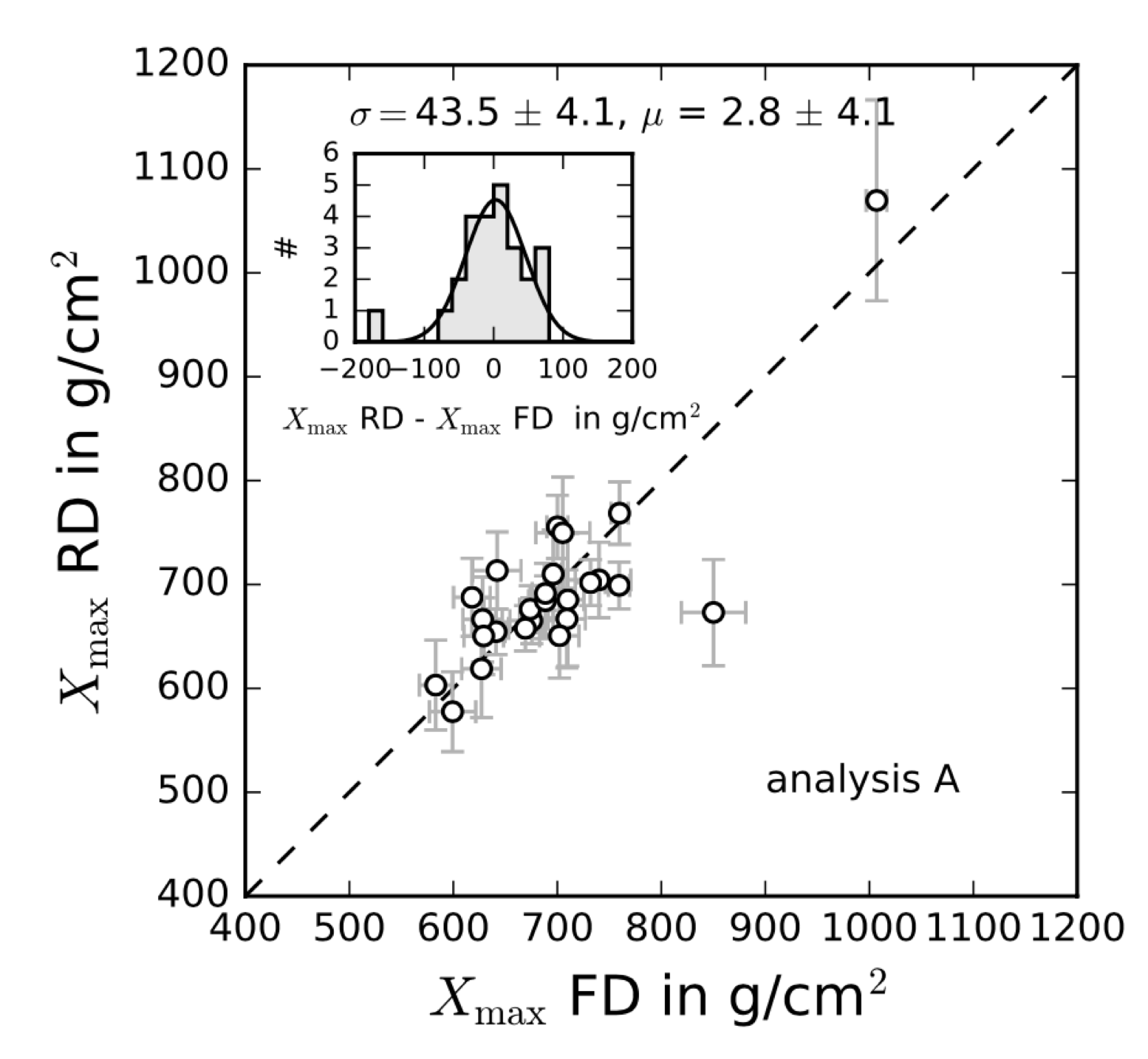}
\caption{{\it Left}, map of the AERA layout within the Pierre Auger Observatory and {\it right}, correlation between the reconstructed $X_{\rm max}$ value from AERA measurements and the one determined with fluorescence detectors, extracted  from~\protect\cite{Huege_2019}.}\label{fig:AERA} 
\end{figure*}

\vspace{0.2cm}
\noindent \textbf{AERA}~\cite{Glaser_2017,Huege_2019,Gottowik:2019vP} the Auger Engineering Radio Array, is an extension of the Pierre Auger Observatory. It started in 2011 at Malargue and aimed at detecting air-showers above $10^{17}$\,eV in the $30-80$\,MHz band. It used a hybrid approach, combining $\sim 150$ of Auger's $1627$ surface detectors ($1600$ water tanks and $27$ fluorescence detectors) with $150$ radio antennas (LPDA and Butterfly) over $17$\,km$^{2}$ as illustrated in the left panel of Fig.~\ref{fig:AERA}. This setup with different types of detectors and antennas allowed to test different experimental designs and detection techniques but also different triggering schemes, combining an external trigger from the surface detectors and self-triggering with FPGAs~\cite{Glaser_2017}. Hence it served as a pathfinder to determine the design of larger experiments. Its hybrid approach also allowed to cross-calibrate the energy measurement of the different detectors to test the radio technique and to develop and test different reconstruction methods. Hence, AERA demonstrated that an energy reconstruction resolution of $17\%$ and an $X_{\rm max}$ reconstruction resolution between $35-40$\,g.cm$^{-2}$ were achievable with radio-detection as shown in the right panel of Fig.~\ref{fig:AERA}. Such performances showed that radio detection was a competitive technique providing comparable results with fluorescence and surface detectors. AERA hence probed the efficiency of radio-detection and opened the path towards next generation large-scale experiments.

\subsection{In ice detection: ARA and ARIANNA}

The Askaryan Radio Array~\cite{Miller_2012,Allison_2016} (ARA)  and the Antarctic Ross Ice Shelf Antenna Neutrino Array~\cite{Arianna,Anker_2019} (ARIANNA) experiments are the direct extension of the AURA and RICE concepts of deep-ice and subsurface radio arrays. Thanks to the validation of the in-ice radio technique, this second generation of experiments tried to scale up in size to reach the UHE neutrino landscape.

\vspace{0.2cm}
\noindent \textbf{ARA}
is operating since 2010 near the Amundsen-Scott station at the South Pole and is the evolution of AURA.
Each station is situated at $\sim200$\,m depth, and made of $4$ strings composed of $16$ antennas operating in the $200-850$\,MHz band, for a volume of $20\times20\times20$\,m$^{3}$.
The $37$ stations currently deployed, in an hexagon with a $2$\,km spacing, lead to an effective volume of $200$\,km$^{3}$ at $1$\,EeV. This volume while being the largest monitored by in-ice radio antennas remains well below the required values since it would need $4$ times more in order to reach the UHE fluxes.

\begin{figure}[tb]
\centering 
\includegraphics[width=0.75\columnwidth]{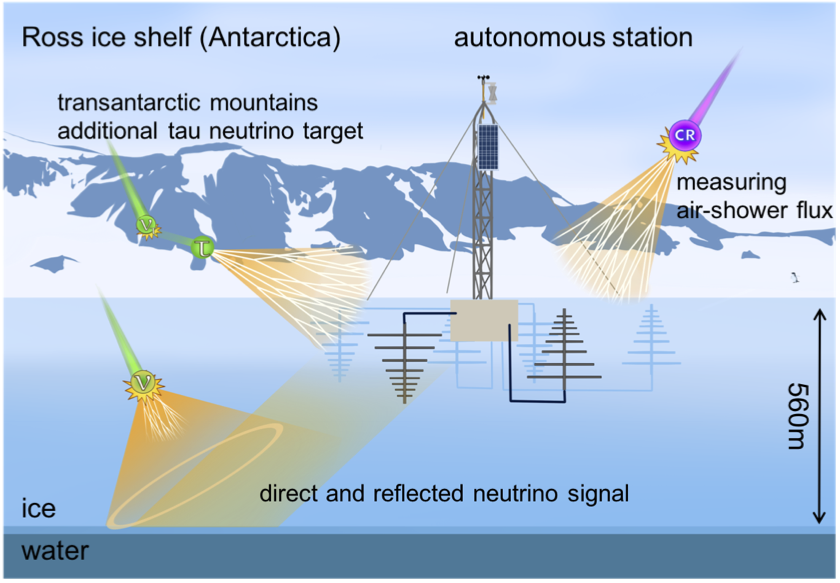}
\caption{Sketch of ARIANNA detection principle at the Ross Ice Shelf from~\protect\cite{Anker_2019}. Neutrino can be detected from air-induced showers after interacting in trans-antarctic mountains or from in-ice showers using the reflected signal at the water-ice interface. Cosmic-ray air showers are detected with upward looking antennas.}\label{fig:ARIANNA}
\end{figure}

\vspace{0.2cm}
\noindent \textbf{ARIANNA}
is located in Moore's bay at $110$\,km from Mc Murdo station, and was in operation between between $2010-2015$.
The initial detector concept is presented in Fig.~\ref{fig:ARIANNA} and relies on a layout of $36\times36$ antenna stations with a $1$\,km spacing. Each station is made of 8 down-looking and 2 upward looking antennas ( for calibration and cosmic-ray veto) operating between $100-1300$\,MHz.
The detector uses radio reflections from the water-ice interface at the bottom of the ice shelf to detect direct and indirect signals, leading to almost a factor $2$ of increase in the field of view and effective volume.
However since the antennas are deployed close to the upper layers of the ice (the firn) the attenuation length is only about $400-500$\,m.
Yet, only a few dozen of stations were deployed, very far from the $1000$ needed to reach the UHE cosmogenic neutrino fluxes, as their concept study required.\\

Initially designed to scale up to reach a thousand cubic kilometers of effective volume in order to be sensitive enough to cosmogenic UHE neutrinos, these two experiments faced various deployment issues and increase of costs, stopping their ambitious goals.
Nevertheless, ARA and ARIANNA have shown that scaling up is possible while maintaining noises at reasonable levels.

\subsection{Balloon detection: ANITA}
\begin{figure}[tb]
\centering 
\includegraphics[width=0.99\columnwidth]{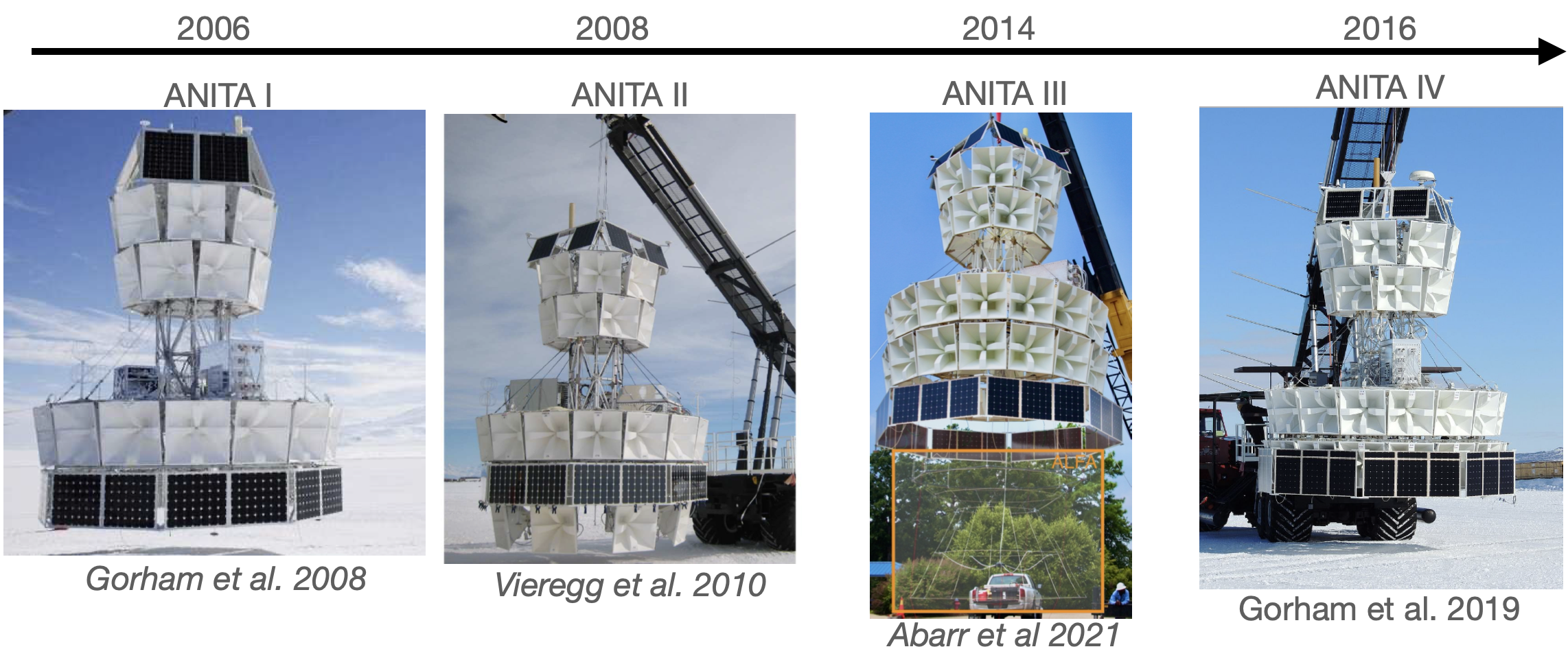}
\caption{Evolution of ANITA payloads over the years.}\label{fig:ANITA}
\end{figure}

\textbf{ANITA,} the ANtarctic Impulsive Transient Antenna~\cite{Gorham_2009,Gorham_2019,Gorham_2021}, is a stratospheric balloon concept that intended to detect both in-ice and in-air showers (see Fig.~\ref{fig:ANITA}). The instrument was designed to be sensitive to geomagnetic (in air) and Askaryan (in ice) emissions, for both down-going and up-going trajectories, and direct and reflected signals.
By flying above Antarctica the successive 5 missions took advantage of the South Pole wind vortex, and the phenomenal amount of ice contained in the south ice cap, allowing to monitor a large volume of ice thanks to the stratospheric altitude (on average $40$\,km) and the long duration flight.
The 5 past missions based on this concept, successively improved and build up on the experience of each previous mission, leading to the best neutrino limits above $100$\,EeV, as well as the best UHE neutrinos candidates, and a few anomalous events yet to explain.

The design of the last mission, called ANITA IV, highlights best the improvements made so far for this kind of missions.
The radio instrument is made of 48 quad ridged horn antennas with high gain and dual polarisation (Horizontal and Vertical), operating in the $180-1200$\,MHz band. The array layout is designed on $3$ cylindrical layers, covering the complete azimuth range in 16 sectors. Finally the RF signal chain allowed for a "threshold riding" trigger system constantly adjusting to the background with an event rate up to $50$\,Hz.
The flight lasted for $\sim30$\,days, during which
2 types of anomalous events were found with very steep trajectories ($\sim-30^{\circ}$) below the horizon, and close to the horizon. They have a non-inverted polarity as expected for neutrino-induced EAS but reconstruction have shown a compatibility with the background at the $\sim3\sigma$ confidence level. The exact nature of these events remains elusive but motivates the development of a new instrument, PUEO, with an improved sensitivity, and able to disentangle with precision the different possible trajectory of events.

\section{Next generation of radio experiments}
The results achieved by the second generation of radio experiments created a momentum in the field towards the challenge of the detection of UHE cosmogenic neutrinos. In particular for the in-air experiments, which, for the first time mostly aim at performing the radio-detection of neutrinos. 
While for the in-ice experiments, the main challenge of this new generation is to solve the scaling issues in order to reach the UHE landscape.

This turning point in the scientific goals pursued by the in-air experiments has initiated for the first time a competition between the in-air and in-ice techniques.

\subsection{In air detection: Auger Prime, BEACON and GRAND}

\begin{figure*}[tb]
\centering 
\includegraphics[width=0.54\columnwidth]{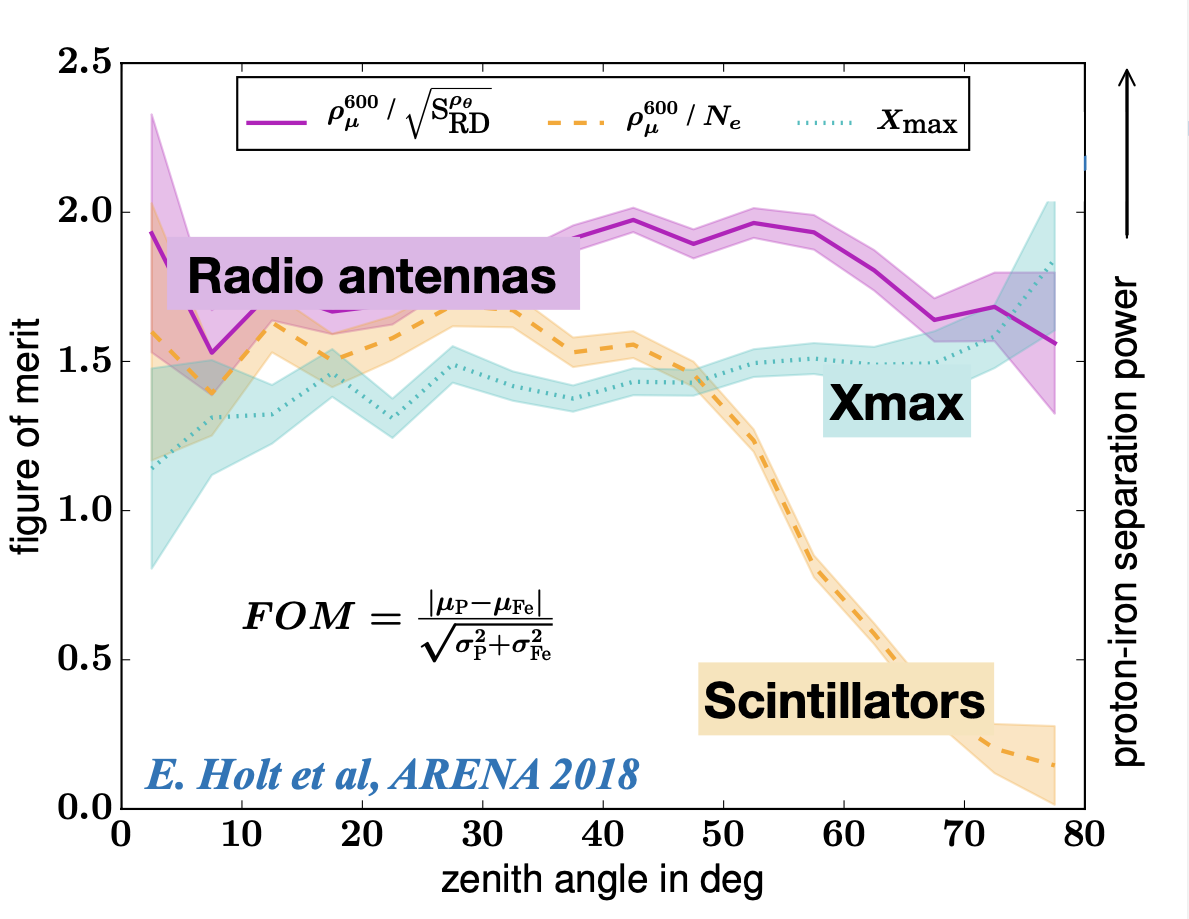}
\includegraphics[width=0.44\columnwidth]{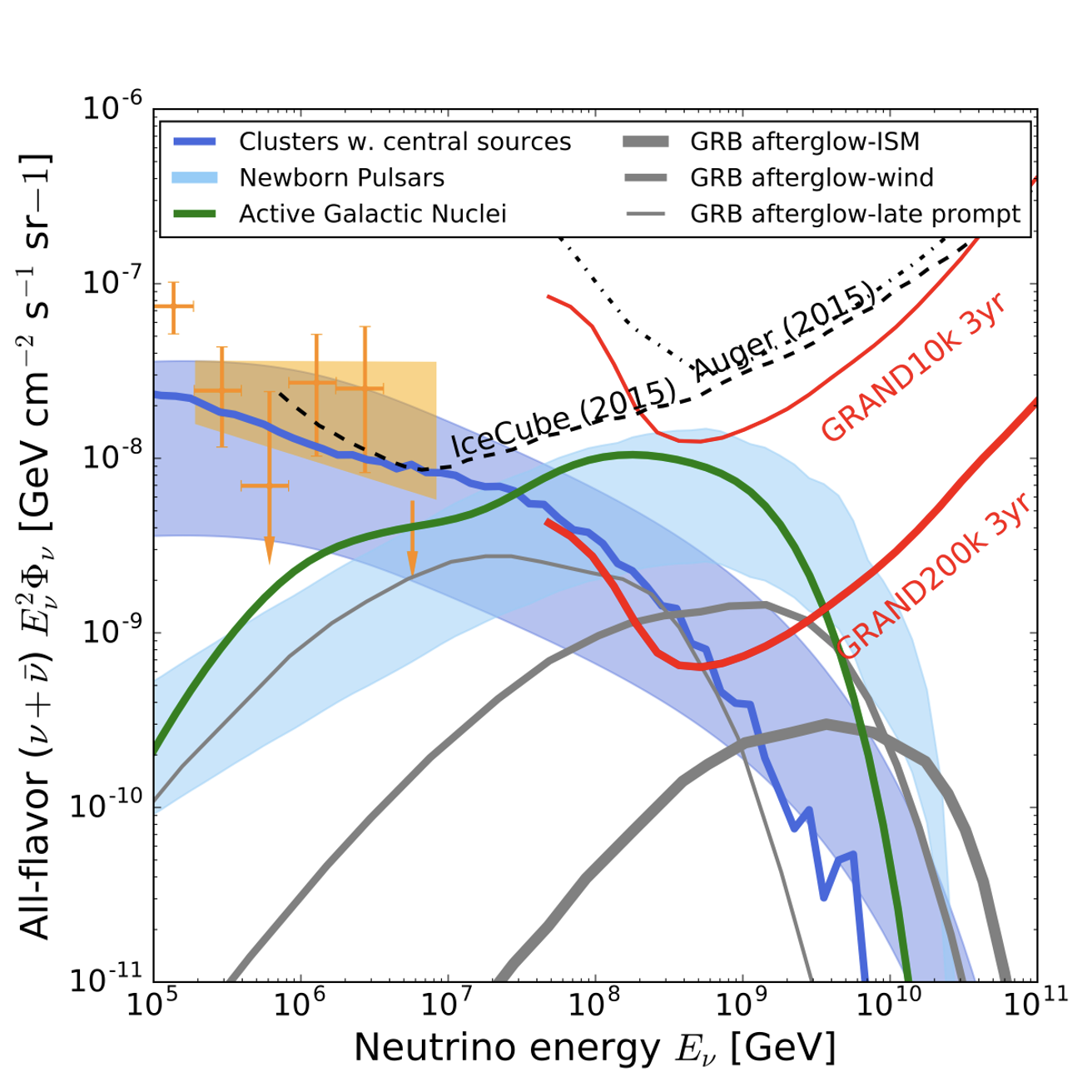}
\caption{{\it Left}, separation power between proton and iron induced air-showers from~\protect\cite{Huege_2019}. {\it Right}, prediction of UHE neutrino fluxes from astrophysical sources compared with upper limits of Auger, IceCube and GRAND. Extracted from~\protect\cite{_lvarez_Mu_iz_2019}.}\label{fig:Next_gen} 
\end{figure*}

\textbf{Auger-Prime}~\cite{Aprime,Stasielak_2022,Castellina_2019} is the upgrade of the Pierre Auger Observatory optimized for electron/muon discrimination. Its main objectives were to (1) study the flux suppression at ultra-high energy to check whether it come from a maximal acceleration energy of the sources or from a GZK cut-off, (2) explain the weak correlation between cosmic-rays and nearby galaxies and (3) put constraints on the hadronic interaction models at ultra-high energy. This translated into a need for an increase of the mass composition sensitivity above $1 \,$EeV. 
Electron/muon discrimination is of particular interest to tackle this objective. As shown in the left panel of Fig.~\ref{fig:Next_gen}, the ratio between the muon content at a lateral distance of 600 meters, $\rho_{\mu}^{600}$, and the number of electrons $N_{e}$ gives a good mass composition sensitivity for vertical showers but breaks for inclined showers where the scintillators efficiency drop considerably. Yet, combining measurements from the radiation energy $S_{\rm RD}^{\rho_{\theta}}$ and  $\rho_{\mu}^{600}$ provide a good sensitivity on the mass composition for any zenith angle. Since inclined showers propagate over large distances, electrons in these showers rapidly die out while muons can reach the ground. Hence inclined showers allow for an efficient discrimination between the muon and the electromagnetic content if radio antennas are combined with particle detectors. As a consequence, AugerPrime aims at adding scintillators and radio antennas for each of the pre-existing surface detectors which should provide a sensitivity on the mass composition comparable or even better to methods using  $X_{\rm max}$ as main observable.

\vspace{0.2cm}
\noindent \textbf{GRAND}~\cite{_lvarez_Mu_iz_2019} the Giant Radio array for Neutrino Detection, is a planned large-scale radio experiment dedicated to the in-air detection of ultra-high energy messengers with a main focus on UHE neutrinos. It should consist of a radio array of $\sim 200\,000$ antennas over $200\,000$\,km$^{2}$ deployed in several mountainous regions regions around the world. The focus is put on very inclined air-showers to detect tau-neutrinos with Earth-skimming trajectories that go through a dense medium as a mountain or Earth's surface for up-going trajectories. The deployment of the GRAND experiment is expected to be staged, i.e., divided in 3 main steps, GRANDproto300, GRAND10k and GRAND200k. GRANDProto300 is the deployment of the first 300 antennas over $\sim 200$\,km$^{2}$ in western China, to detect cosmic-rays and hopefully gamma-rays in the $10^{16.5}-10^{18}$\,eV energy range. It will serve as a test bench for the GRAND experiment, validating the detection and the reconstruction feasibility of highly inclined showers ($\theta>80^{\circ}$) by probing autonomous radio-detection on large-scale arrays and angular reconstruction below $0.1^{\circ}$. To tackle these objectives, the radio antennas may be combined with particle detectors. This first layout will be followed by GRAND10k an extension of GRANDProto300 to $\sim\,10\,000$ antennas expected for $\sim 2025$ and allowing us to detect the first UHE neutrinos for optimistic fluxes. Finally GRAND200k will consist of 20 sub-arrays of 10 000 antennas all around the world. It should reach a tau-neutrino sensitivity of $4\times10^{-10}E^{-2}$\,GeV\, cm$^{-2}$\,s$^{-1}$\,sr$^{-1}$ within 3 years of operation as seen in the right panel of Fig.~\ref{fig:Next_gen} and $4\times10^{-10}E^{-2}$\,GeV\, cm$^{-2}$\,s$^{-1}$\,sr$^{-1}$ within 10 years, surely allowing for the detection of UHE neutrinos and opening the path towards UHE neutrino astronomy.

\vspace{0.2cm}
\noindent \textbf{BEACON}~\cite{Wissel_2020,Southall:2021mc}
The Beam forming Elevated Array for COsmic Neutrinos is a detector that envisions to use the radio interferometry technique, in the $30-80$\,MHz range to detect Earth-skimming tau neutrino-induced events. The concept aims at deploying antenna stations atop of high elevation mountains in order to increase the field of view towards the ground, hence increasing the collecting area of radio signals from events emerging below the horizon and propagating in upward trajectories.
This project is therefore building on two key elements: first the radio-interferometer technique, extensively used by the radio-astronomy community for high sensitivity observations, and second its topography site, which provide a large field of view. Preliminary simulation studies have shown that in principle BEACON could reach down to $6.10^{-9} \rm GeV \, cm^{-2}\, s^{-1} \, sr^{-1}$ at neutrino energies of $\sim 1$\,EeV, after 3 years of integration and for a thousand antenna stations.

Currently a prototype of $8$ antennas is deployed at the Barcroft Station in the White Mountains of California. The prototype, too small to envision significant neutrinos detection for the moment, is used as a test-bench for calibration and data analysis on cosmic-ray observations. Simulation studies have shown that tens of events are already detectable with its current configuration. Once the validation of this prototype will be achieved the up-scaling of the experiment can be easily achieved by the deployment of a few hundreds of stations making it ultimately competitive to detect UHE neutrinos.

\subsection{In ice detection: RNO-G and IceCube-Gen2}

\begin{figure}[tb]
\centering 
\includegraphics[width=0.99\columnwidth]{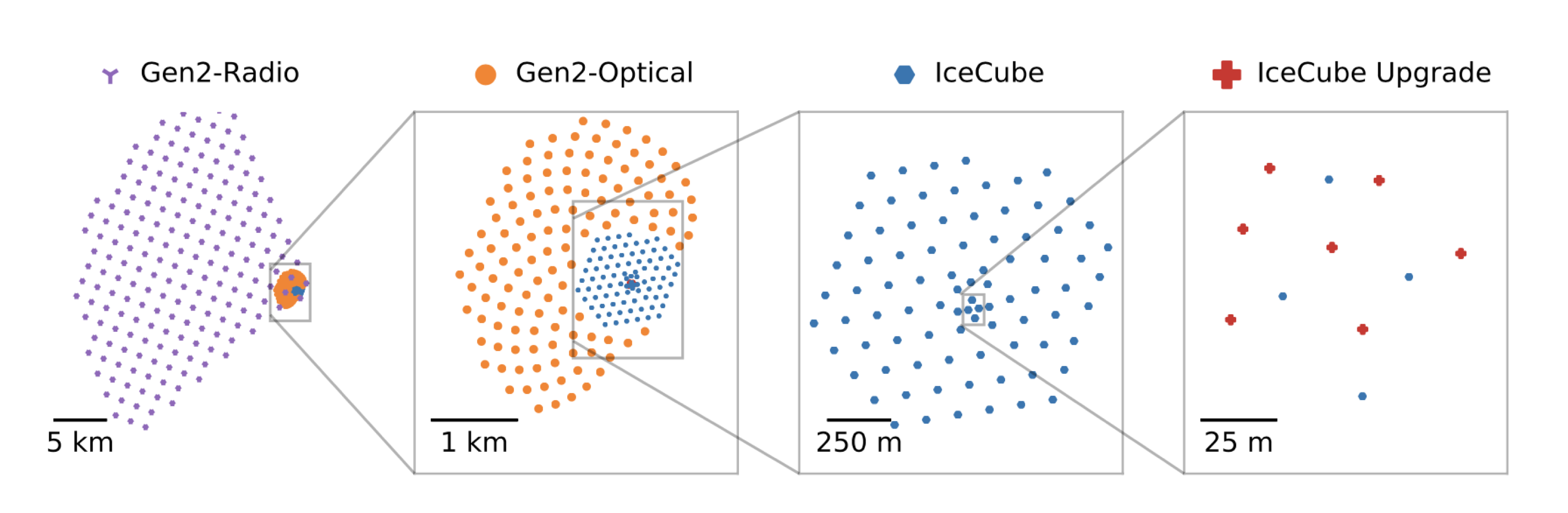}
\caption{Planned design of the IceCube-Gen2 Neutrino Observatory from~\protect\cite{Aartsen_2021}}\label{fig:IceCube}
\end{figure}

\textbf{IceCube}~\cite{Clark_2021,Aartsen_2021,Icube} is an in-ice experiment initiated in 2010 for high energy astrophysical neutrino detection in the Antarctic. It combines an in-ice detector “IceCube”, with a surface array “Ice Top”. The ice detector consists of several strings regrouping a total 5160 Digital Optical Modules made of photomultiplicators that detect the Cerenkov light from neutrino-induced cascades in ice. IceCube can detect two types of events, tracks events, from charged current interaction of muon neutrinos and cascade events from either charged current interactions of electron and muon neutrinos or neutral current interaction of all three flavors. Track events allow for a better angular resolution while cascades events are better for the energy reconstruction. Since it started IceCube detected neutrinos between 10 TeV and 10 PeV and even allowed for the identification of a point source at the 3$\sigma$ confidence level, the Blazar TXS0506+056 in 2017.
IceCube-Gen2 is the planned upgrade of the IceCube observatory. It will be $\sim$ 20 times larger and will aim at (1) increasing the rate of observed cosmic neutrino events by an order of magnitude, (2) broadening the energy range from TeV-PeV to GeV-EeV and (3) increasing the angular resolution to identify more point sources. For this purpose, 4 additional components will be added to IceCube as illustrated in Fig.~\ref{fig:IceCube}:

\begin{itemize}
    \itemsep0em 
    \item IceCube Upgrade, 7 additional strings denser in Digital Optical Modules (693 sensors in total, 99 for each string) to lower the low energy threshold.
    \item Gen2-Optical, an enlarged optical array with 120 new string (with 60 DOMs, 17 meters apart for each string)  in a sunflower pattern with 240 meters of lateral spacing to increase the sensitivity to TeV-PeV neutrinos
    \item Gen2-Surface, additional radio antennas and scintillators co-located with each of the Gen2-optical new string to perform  cosmic-ray detection at PeV energies. This setup should also help as a veto for down-going cosmic-ray showers.
    \item Gen2-Radio, $\sim$ 200 clusters of antennas over 500$\, {\rm km^{2}}$ for the detection of EeV neutrinos. This setup will combine antennas at the surface and deeper with a phased-array system to beam-form the signal and lower the trigger thresholds.
\end{itemize}

\vspace{0.2cm}
\noindent \textbf{RNO-G}~\cite{Aguilar_2021}
The Radio Neutrino Observatory in Greenland is located at the Summit Station (Greenland), and will target neutrinos from several $\rm PeV$ to the $\rm EeV$ range. It relies on ARA and ARIANNA detector concepts, combining the advantages of both detection principles, while minimizing the number of antenna stations required to reach the effective volume needed to detect the UHE cosmogenic neutrinos.
Each station is composed of a deep-ice log-periodic antenna array ($150-600$\,MHz), {\it a la ARA}, designed with a phased triggering system, and a set of subsurface antennas ($100-1300$\,MHz), a la ARIANNA, oriented in order to be able to fully measure the polarisation of any shower induced signal, hence increasing the reconstruction capabilities of the station.
The deep-ice array is made of $3$ strings plunging into the ice sheet, down to $100$\,m and delimiting a monitored ice volume of roughly $1$\,km$^3$. 
The envisioned configuration is $35$ independent stations, in order to reach the UHE cosmogenic neutrino flux. The design of RNO-G will serve as a reference to build the IceCube-Gen2 radio detector.

\subsection{Balloon detection: PUEO}
\textbf{PUEO}~\cite{Abarr_2021} the Payload for Ultrahigh Energy Observations is the direct successor of the ANITA program. 
The challenge envisioned by this mission is to increase by almost two orders of magnitude the combined sensitivity of all previous ANITA missions (from ANITA I to ANITA IV), while keeping the same constraints in term of detector size (because of stratospheric balloon requirements). To tackle this challenge, PUEO can build on the experience of the past ANITA missions, as well as the recent developments made in the radio field, especially towards hardware and firmware improvements.
Indeed, this new payload is composed of two instruments:
\begin{itemize}
\itemsep0em 
\item the Main Instrument (MI) is composed of 108 quad-ridged horn antennas  (twice as much as the last ANITA mission), canted at $-10^{\circ}$, and targeting the $300-1200$\,MHz band across two polarisation channels (Horizontal --H and Vertical --V), as well as $12$ antennas canted at $-40^{\circ}$ in the direction of the ground, and dedicated to disentangle the very steep trajectories. The main instrument, hence, totalizes $240$ channels organized in $4$  cylindrical layers of $24$ azimuthal sectors. Furthermore it is using a phased array trigger technologies, allowing for a drastic reduction of the trigger threshold SNR, down to $1-2$, while maintaining a high detection efficiency.
\item the Low Frequency Instrument (LFI) is composed of 8 sinuous antennas observing in the $50-300$\,MHz band, across two polarisation (H and V). This instrument is dedicated to the detection of the Earth-skimming tau neutrino-induced EAS, and while having only 16 channels, its effective area is twice larger than the MI, towards these events. The explanation for  such performances comes from the large collecting aperture of each antennas (about $1.9$\,m in diameter) but more importantly from the chosen frequency band, where the radio beam induced by EAS is larger by almost a factor $2$ compared to the MI frequency band.
\end{itemize}
The decision to split the frequency band between two instruments allows to more than double the number of channels, up to 256 in total, for the same volume of payload, while restricting the radio noise, more intense in the low frequency band, to only one instrument.

Thanks to all these improvements and new developments, PUEO should be able to reach sensitivities down to $\sim3.10^{-9}$\,GeV cm$^{-2}$ s$^{-1}$ sr$^{-1}$ at neutrino energies of $\sim 10$\,EeV, from simulation studies.
PUEO is envision to fly in the summer 2024.

\section{Summary}
\begin{figure}[tb]
\centering 
\includegraphics[width=0.90\columnwidth]{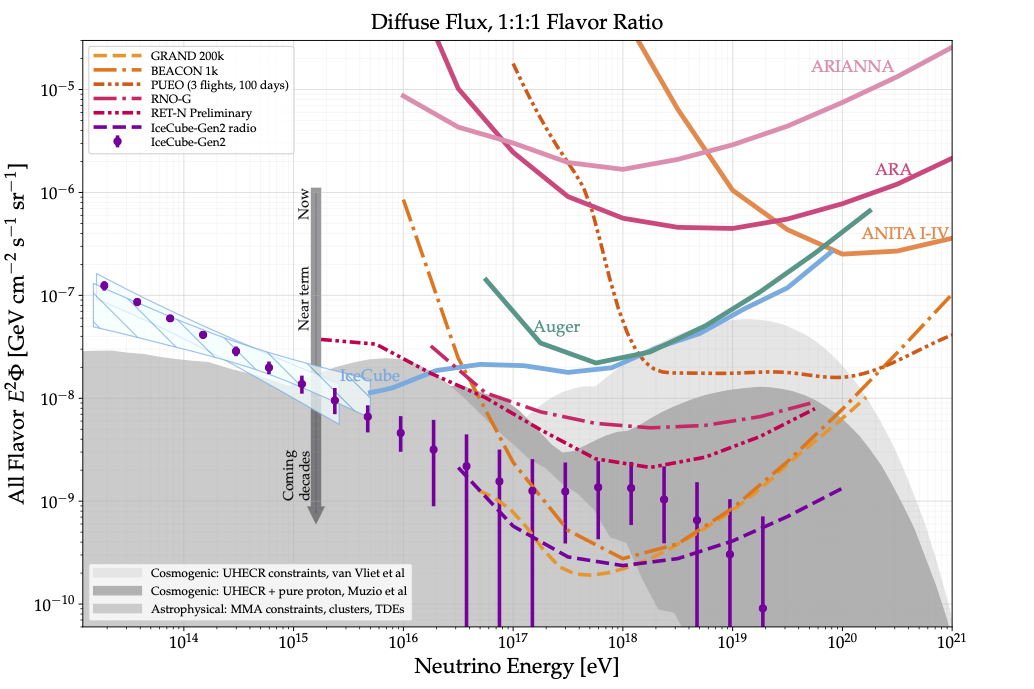}
\caption{Expected differential 90\% CL sensitivity to a diffuse neutrino
flux adapted from~\protect\cite{roshan_2022} by Stephanie Wissel. A 10 years exposure is assumed for planned experiments unless otherwise noted. The blue band indicates the astrophysical measured neutrino flux by IceCube and the solid lines the experimental upper limits for high energy neutrinos.}\label{fig:nu_sensitivity}
\end{figure}

Over the past 2 decades, major improvements for radio-detection of astroparticles were achieved. This was motivated by an expected duty cycle of 100$\%$ combined with the low costs of radio antennas and allowed by the development of powerful digital signal processing techniques. First progresses were possible thanks to the pioneering experiments of CODALEMA, LOPES or RICE and AURA that probed respectively that in-air and in-ice detection were feasible. This then led to the second generation of experiments with LOFAR, TREND and AERA for in air, ARA and ARIANNA for in ice, and ANITA for both in air and in ice that confirmed the feasibility of radio-detection but also showed that it could be competitive with already existing detection methods using surface detectors. These experiments also provided a better understanding of the radio emission processes, allowed to test different experimental designs and to develop efficient reconstruction methods. AERA and TREND in particular probed the feasibility of autonomous radio detection and opened the path towards future large-scale radio experiments. Radio-detection is now established as a cheap, reliable and competitive technique to detect astroparticles and the objective is now focus towards building increasingly larger facilities to assess the low fluxes at ultra-high energy. Towards this goal, several experiments are already planned or in construction as GRAND, AugerPrime, BEACON and PUEO for in air-detection and IceCube-Gen2, RNO-G and PUEO for in-ice detection and other experiments with innovative detection methods should follow with RET~\cite{Prohira_2021}, the Radar Echo Telescope and GCOS~\cite{Hrandel_2021}, the Global Cosmic Ray Observatory.  These experiments now aim at combining hybrid detection methods and/or a multi-messenger approach and should reach an unprecedented sensitivity to UHE cosmic-rays, gamma-rays and particularly to UHE neutrinos as displayed in Fig.~\ref{fig:nu_sensitivity}. All these efforts toward radio-detection are promising to detect the first UHE neutrinos and establish candidate UHECRs sources by the 2030s.

\section*{Acknowledgments}

The authors want to acknowledge Oliver Martineau-Huynh for the fruitful discussions and inputs on GRAND and other experiments, Jakob Henrichs for his precious comments. We also want to thank the Moriond conference organizing committee and particularly Vera for setting the event and their kind invitation.
VD wishes to thank Stephanie Wissel, Yuchieh Ku and Andrew Zeolla for their inputs and discussions on the experiments of PUEO, BEACON, ARA and RNO-G.
This work is supported by the APACHE ANR grant (ANR-16-CE31- 0001) of the French Agence Nationale de la Recherche, the Programme National des Hautes Energies of CNRS/INSU with INP and IN2P3, co-funded by CEA and CNES. VD's work is founded by the NASA APRA and Pioneer programs (80NSSC20K0925 and 80NSSC21M0116), and the NSF CAREER program (2033500).

%\bibliography{tempbib}

%\subsection{Photograph}

%You may want to include a photograph of yourself below the title
%of your talk. A scanned photo can be 
%directly included using the default command\\
%\verb^\newcommand{\Photo}{\includegraphics[height=35mm]{mypicture}}^\\
%just before the 
%\verb^\begin{document}^
%line. If you don't want to include your photo, just comment this line
%by adding the \verb^%^ sign at the beginning of 
%the line and uncomment the next one
%\verb^%\newcommand{\Photo}{}^ by removing its \verb^%^ sign.

\end{document}